\documentclass[12pt]{article}%
\usepackage{amsmath,latexsym}
\usepackage{graphicx}
\usepackage{amsmath}
\usepackage{amsfonts}
\usepackage{amssymb}%
\begin{document}

\title{{Connecting noncommutative geometry to $f(R)$
   modified gravity}}
   \author{
Peter K  F Kuhfittig*\\
\footnote{E-mail: kuhfitti@msoe.edu}
 \small Department of Mathematics, Milwaukee School of
Engineering,\\
\small Milwaukee, Wisconsin 53202-3109, USA}

\date{}
 \maketitle

\begin{abstract}\noindent
It is shown in this note that a
noncommutative-geometry background determines
the modified-gravity function $f(R)$ for
modeling dark matter. \\


\end{abstract}

\section{Introduction}

It is well known that $f(R)$ modified
gravitational theories can account for dark
matter in the sense that the galactic dynamics
of massive test particles can be explained
in the framework of $f(R)$ gravity without
the need for dark matter \cite{CCT, BGS, MS}.
For a general discussion of dark matter as a
geometric effect of $f(R)$ gravity, see Ref.
\cite{BTH}.  Less well known is that
noncommutative geometry can play a similar
role \cite{pK17, KG14}.  The purpose of this
note is to explain the reason for the
connection between the two theories.  Given
that noncommutative geometry is an offshoot
of string theory, the results of this note
may be viewed as indirect evidence for the
latter.

\section{Noncommutative geometry and $f(R)$
    modified gravity}

Noncommutative geometry is based on the
following outcome of string theory:
coordinates may become noncommuting
operators on a $D$-brane \cite {eW96, SW99}.
This statement refers to the commutator
$[\textbf{x}^{\mu},\textbf{x}^{\nu}]=
i\theta^{\mu\nu}$, where $\theta^{\mu\nu}$
is an antisymmetric matrix.  Noncommutativity
replaces point-like structures by smeared
objects.  As discussed in Refs.
\cite{SS1, SS2}, the aim is to eliminate
the divergences that normally occur in
general relativity.  A good way to
accomplish the smearing effect is to
assume that the energy density of the
static and spherically symmetric and
particle-like gravitational source has
the form \cite{NM08, pK15}
\begin{equation}\label{E:rho1}
  \rho(r)=\frac{M\sqrt{\beta}}
     {\pi^2(r^2+\beta)^2}
\end{equation}
in spherical coordinates and is interpreted
to mean that the mass $M$ of the particle
is diffused throughout the region of linear
dimension $\sqrt{\beta}$ due to the
uncertainty.

The study of dark matter and dark energy has
led to a renewed interest in modified theories
of gravity.  One of the most important of
these, $f(R)$ modified gravity, replaces the
Ricci scalar $R$ in the Einstein-Hilbert
action
\begin{equation*}
  S_{\text{EH}}=\int\sqrt{-g}\,R\,d^4x
\end{equation*}
by a nonlinear function $f(R)$:
\begin{equation*}
   S_{f(R)}=\int\sqrt{-g}\,f(R)\,d^4x.
\end{equation*}
(For a review, see Refs.
\cite{NO07, fL08, SF10}.)

Since both theories can account for dark
matter, it is important to determine a
possible connection.

\section{Noncommutative geometry and dark
   matter}

We start with the general metric of a static
and spherically symmetric line element,
using units in which $c=G=1$ \cite{MTW}:
\begin{equation}\label{E:line}
ds^{2}= -e^{2\Phi(r)}dt^{2}+
\frac{dr^2}{1-\frac{2m(r)}{r}}
+r^{2}(d\theta^{2}+\text{sin}^{2}
\theta\,d\phi^{2}).
\end{equation}
Here $m(r)$ is the effective mass inside a
sphere of radius $r$ with $m(0)=0$.  We
also require that
$\text{lim}_{r\rightarrow\infty}\Phi(r)=0$
and
$\text{lim}_{r\rightarrow\infty} m(r)/r=0$,
usually referred to as asymptotic flatness.

Because of the spherical symmetry, the only
nonzero components of the stress-energy
tensor are $T^t_{\phantom{tt}t}=-\rho(r)$,
the energy density, $T^r_{\phantom{rr}r}=
p_r(r)$, the radial pressure, and
$T^\theta_{\phantom{\theta\theta}\theta}=
T^\phi_{\phantom{\phi\phi}\phi}=p_t(r)$,
the lateral pressure.  Assuming the
conservation law $T^{\alpha}
_{\phantom{\beta r}\beta;\,\alpha}=0$, there
are only two independent Einstein field
equations,
\begin{equation}\label{E:Einstein1}
   \rho(r)=\frac{1}{8\pi}\frac{2m'(r)}{r^2}
\end{equation}
and
\begin{equation}\label{E:Einstein2}
   p_r(r)=\frac{1}{8\pi}\left[-\frac{2m(r)}{r^3}
   +\frac{2\Phi'(r)}{r}\left(1-\frac{2m(r)}{r}
   \right)\right].
\end{equation}

Next, we need to recall that galaxies exhibit
flat rotation curves (constant tangential
velocities) sufficiently far from the
galactic center, due to the existence of
dark matter \cite{RTF80}.  This behavior
indicates that the matter in the galaxy
increases linearly in the outward radial
direction.  More precisely, the total mass
$M_T(r)$ enclosed in a sphere of radius
$r$ has the form
\begin{equation}\label{E:tangential1}
   M_T(r)=v^2r,
\end{equation}
where $v$ is the constant tangential
velocity; here $v^2=0.000001$ in
geometrized units \cite{Nandi}.

To connect these ideas to noncommutative
geometry, we start with a thin spherical
shell of radius $r=r_0$.  So instead of
a smeared object, we now have a smeared
spherical surface.  Let us consider the
smearing in the outward radial direction
only, that being the analogue of a
smeared particle at the origin.  So
Eq. (\ref{E:rho1}) is replaced by
\begin{equation}\label{E:rho2}
    \rho(r)=\frac{M_{r_0}\sqrt{\beta}}
    {\pi^2[(r-r_0)^2+\beta]^2},
\end{equation}
where $M_{r_0}$ is the mass of the shell.
According to Ref. \cite{pK17}, the smeared
mass $m_{\beta}(r)$ is given by
\begin{equation}\label{E:m(r)}
   m_{\beta}(r)=\\ \frac{2M_{r_0}}{\pi}
   \left[\text{tan}^{-1}\frac{r-r_0}
   {\sqrt{\beta}}-\frac{(r-r_0)\sqrt{\beta}}
   {(r-r_0)^2+\beta}\right].
\end{equation}
Observe that
\[
   \text{lim}_{\beta\rightarrow 0}\,m_{\beta}
   (r)=M_{r_0}.
\]
So the mass of the shell is zero at $r=r_0$
and rapidly rises to $M_{r_0}$.  (We will see
later that $r-r_0$ has to exceed $\sqrt
{\beta}$.)

It is also shown in Ref. \cite{pK17} that
\begin{equation}\label{E:tangential2}
  M_T(r)\approx M_{r_0}(r-r_0),
\end{equation}
in agreement with Eq. (\ref{E:tangential1}).
As noted in Ref. \cite{pK17}, $M_{r_0}$ must
now be viewed as a dimensionless constant of
proportionality which can be interpreted as
the change in the smeared mass per unit length
and is therefore constant throughout.  This
also follows from Eq. (\ref{E:tangential2})
since $dM_T(r)/dr=M_{r_0}$.  We also have
from Eqs. (\ref{E:tangential1}) and
(\ref{E:tangential2}) that $v^2\approx 
M_{r_0}\left(1-\frac{r_0}{r}\right)$.  So
for reasonably large $r$,
\begin{equation}\label{E:v}
   v^2\approx M_{r_0}.
\end{equation}
To reiterate, $v^2$ is approximately equal
to the change in the smeared mass per unit
length.  Since Eq. (\ref{E:v}) holds for
every shell, we could simply replace
$r-r_0$ in Eq. (\ref{E:rho2}) by a new
variable.  However, from a calculational
standpoint, it would be simpler to let
$r_0=0$.  Then Eq. (\ref{E:rho2}) becomes
\begin{equation}\label{E:newrho}
  \rho(r)=\frac{M_{r_0}\sqrt{\beta}}
     {\pi^2(r^2+\beta)^2},
\end{equation}
where $M_{r_0}$ now assumes its original
meaning as the mass of the shell.

As a final comment, it was noted after
Eq. (\ref{E:m(r)}) that $r-r_0$ must
exceed $\sqrt{\beta}$.  To make use of
Eq. (\ref{E:newrho}) in Sec.
\ref{S:connection}, we will need the
more precise condition (with $r_0=0$)
\begin{equation}\label{E:condition}
   r\geq a>\sqrt{\beta}, \quad a>0.
\end{equation}

\section{The connection to $f(R)$ gravity}
\label{S:connection}
Returning now to $f(R)$ gravity, it is
convenient, in view of line element
(\ref{E:line}), to denote $M_T(r)$ by
$m(r)$.  According to Ref. \cite{LO09},
the Ricci scalar $R$ is given by
\begin{equation}\label{E:Ricci1}
    R=\frac{4m'(r)}{r^2}.
\end{equation}
From Eq. (\ref{E:tangential1}) we then
obtain
\begin{equation}
   \frac{dM_T(r)}{dr}=m'(r)=v^2
\end{equation}
and
\begin{equation}\label{E:Ricci2}
   R(r)=\frac{4v^2}{r^2}.
\end{equation}
This equation yields
\begin{equation}\label{E:rofR}
   r(R)=\sqrt{\frac{4v^2}{R}}.
\end{equation}
In $f(R)$ gravity, Eq. (\ref{E:Einstein1})
is replaced by \cite{LO09}
\begin{equation}\label{E:Einstein3}
   \rho(r)=F(r)\frac{2m'(r)}{r^2},
\end{equation}
where $F=\frac{df}{dR}$.  Eq.
(\ref{E:newrho}) now yields
\begin{equation}
   F(r)=\frac{r^2}{2m'(r)}\rho(r)
   =\frac{r^2}{2m'(r)}\frac{M_{r_0}
   \sqrt{\beta}}{\pi^2(r^2+\beta)^2}
\end{equation}
and from Eqs. (\ref{E:Ricci1}) and
(\ref{E:rofR}),
\begin{equation}
   F(R)=\frac{2M_{r_0}\sqrt{\beta}}{\pi^2}
   \frac{1}{R\left(\frac{4v^2}{R}+\beta
   \right)^2}.
\end{equation}
Integrating, we get
\begin{multline}\label{E:f(R)}
   f(R)=\frac{2M_{r_0}\sqrt{\beta}}
   {\pi^2\beta^2}\left[\text{ln}\,
   (\beta R+4v^2)-\frac{\beta R}
   {\beta R+4v^2}+\text{ln}\,C\right]\\
   =\frac{2M_{r_0}}{\pi^2\beta^{3/2}}
   \left[\text{ln}\,(4v^2)+\text{ln}
   \left(1+\frac{\beta R}{4v^2}\right)
   -\frac{\beta R}{\beta R+4v^2}
   +\text{ln}\,C\right],
\end{multline}
where $C$ is an arbitrary constant.

To simplify the analysis, consider the
third term inside the brackets on the
right-hand side of Eq. (\ref{E:f(R)}).
From Condition (\ref{E:condition}),
$r\geq a>\sqrt{\beta}, a>0$, $r$ is
bounded away from 0.  As a result,
\[
   \left|-\frac{\beta}{\beta+4v^2/R}
   \right|=\left|-\frac{\beta}{\beta+r^2}
   \right|\le \left|-\frac{\beta}
   {\beta+a^2}\right|\approx 0
\]
since $a^2$ is fixed and $\beta$ is close
to zero.

In Eq. (\ref{E:f(R)}), letting $\text{ln}\,C
=-\text{ln}\,4v^2$ results in
\begin{equation}
   f(R)\approx\frac{2M_{r_0}}{\pi^2\beta^{3/2}}
   \,\text{ln}\left(1+\frac{\beta R}{4v^2}
   \right).
\end{equation}
Now recalling that $R=4v^2/r^2$, we have
\[
    \frac{\beta R}{4v^2}=\frac{\beta}{r^2}<1
\]
since $r>\sqrt{\beta}$.  Thus
\begin{equation*}
   \text{ln}\left(1+\frac{\beta R}{4v^2}
   \right)=\frac{\beta R}{4v^2}-
   \frac{1}{2}\left(\frac{\beta R}{4v^2}
   \right)^2+\frac{1}{3}\left(
   \frac{\beta R}{4v^2}\right)^3-
   \cdot\cdot\cdot\,\approx\frac
   {\beta R}{4v^2}
\end{equation*}
and
\begin{equation}\label{E:final1}
   f(R)\approx\frac{M_{r_0}}
   {2v^2\,\pi^2\sqrt{\beta}}R.
\end{equation}
Observe that the coefficient of $R$
is a dimensionless constant.  Eq.
(\ref{E:final1}) also implies that
\begin{equation}\label{E:final2}
   f(R)\approx\frac{M_{r_0}}
   {2v^2\,\pi^2\sqrt{\beta}}
   R^{1+\epsilon},\quad
   \epsilon\ll 1.
\end{equation}

According to Ref. \cite{BTH},
$f(R)$ modified gravity can account
for flat galactic rotation curves if
\begin{equation}
   f(R)=kR^{1+\epsilon}, \quad
   \epsilon \ll 1, \quad\text{with
   constant}\,\,k,
\end{equation}
which has the same form as Eq.
(\ref{E:final2}).

\section{Conclusion}
It is shown in this note that a
noncommutative-geometry background
yields the form
\begin{equation}
   f(R)\approx kR^{1+\epsilon}, \quad
   \epsilon \ll 1, \quad\text{with
   constant}\,\,k,
\end{equation}
which is known to account for
galactic rotation curves.
Noncommutative geometry can
therefore serve as a model for
dark matter.

\end{document}